\documentstyle{article}
\def\lsim{\:\raisebox{-0.5ex}{$\stackrel{\textstyle<}{\sim}$}\:}
\def\gsim{\:\raisebox{-0.5ex}{$\stackrel{\textstyle>}{\sim}$}\:}
\def\21{$SU(2) \otimes U(1)$}
\def\321{$SU(3) \otimes SU(2) \ot U(1)$}

\def\neu{\hbox{neutrino }}

\def\beq{\begin{equation}}
\def\eeq{\end{equation}}
\def\bef{\begin{figure}}
\def\eef{\end{figure}}
\def\bet{\begin{table}}
\def\eet{\end{table}}
\def\bea{\begin{eqnarray}}
\def\ba{\begin{array}}
\def\ea{\end{array}}
\def\bi{\begin{itemize}}
\def\ei{\end{itemize}}
\def\ben{\begin{enumerate}}
\def\een{\end{enumerate}}
\def\ra{\rightarrow}
\def\nt{\hbox{$\nu_\tau$ }}
\def\nm{\hbox{$\nu_\mu$ }}
\def\ne{\hbox{$\nu_e$ }}

\def\lsim{\raise0.3ex\hbox{$\;<$\kern-0.75em\raise-1.1ex\hbox{$\sim\;$}}}
\def\gsim{\raise0.3ex\hbox{$\;>$\kern-0.75em\raise-1.1ex\hbox{$\sim\;$}}}

\def\apj#1#2#3{          {\it Astrophys. J. }{\bf #1} (19#2) #3}

\def\nat#1#2#3{          {\it Nature }{\bf #1} (19#2) #3}

\def\np#1#2#3{           {\it Nucl. Phys. }{\bf #1} (19#2) #3}
\def\pl#1#2#3{           {\it Phys. Lett. }{\bf #1} (19#2) #3}
\def\pr#1#2#3{           {\it Phys. Rev. }{\bf #1} (19#2) #3}

\def\prl#1#2#3{          {\it Phys. Rev. Lett. }{\bf #1} (19#2) #3}

\def\n.c.#1#2#3{         {\it Nuovo Cim. }{\bf #1} (19#2) #3}
\def\r.n.c.#1#2#3{       {\it Riv. del Nuovo Cim. }{\bf #1} (19#2) #3}

\def\yf#1#2#3{           {\it Yad. Fiz. }{\bf #1} (19#2) #3}

\def\ppnp#1#2#3{           {\it Prog. Part. Nucl. Phys. }{\bf #1} (19#2) #3}

\begin{document}
\begin{titlepage}
\today
\begin{center}
\hfill{FTUV/95-14}\\
\hfill{IFIC/95-14}\\
\vskip 0.3cm
{\Large \bf Structure formation with a decaying MeV tau neutrino
and a KeV majoron}\\[.5in]
\vskip 1.0cm
{\large \bf A.D.~Dolgov}
\footnote{Permanent address: ITEP, 113259, Moscow, Russia.},
{\large \bf S.~Pastor and J.W.F.~Valle}
\footnote{E-mail valle@flamenco.ific.uv.es}\\
\vskip .5cm
{\it Instituto de F\'{\i}sica Corpuscular - C.S.I.C.\\
Departament de F\'{\i}sica Te\`orica, Universitat de Val\`encia\\
46100 Burjassot, Val\`encia, SPAIN}\\
\end{center}

\begin{quotation}

We consider the scenario of large scale structure formation with
tau neutrino with mass in the MeV range and lifetime of order
of years which decays into a massive majoron with KeV mass. The latter
are the present-day cold dark matter particles. In contrast to the usual
collisionless dark matter, the majorons have a relatively strong
self-interaction and the picture of structure formation is rather
different from the standard. The decay  $J \ra \gamma \gamma$
leads to the existence of an X-ray line at $E_\gamma = m_J/2$
which could be detectable. Electron and muon neutrinos are expected
to be very light, as required in order to account for the solar
neutrino deficit through \ne to \nm oscillations. Supersymmetry
with spontaneously broken R parity provides
a natural particle physics model for our scenario.

\end{quotation}
\end{titlepage}

The measurement by COBE  \cite{cobe1} of the large angle
anisotropy of the cosmic microwave background radiation
has determined the magnitude of primordial density
fluctuations at very large scales. This has ruled out
the simple cold dark matter (CDM) scenario of structure
formation \cite{cdm}, since with the power on large scales
fixed as measured by COBE and with the assumption of a scale
free primordial fluctuations spectrum the model leads to
approximately twice more power at galactic and cluster
scales than observed. If one wants to describe structure
formation in a model with a flat primordial fluctuation
spectrum one has to assume that the onset of matter dominance
(MD) took place later than in a simple CDM model. This can be
achieved e.g. in an open universe with $h^2 \Omega \approx 0.2$,
where $h = H /(100 km/sec/Mpc)$ and $\Omega = \rho / \rho_c$.
This model is disfavored by inflationary scenarios
which (at least in simple versions) predict $\Omega = 1$.
Another possibility is the mixed (hot+cold) dark matter
scenario \cite{cdm} with $\Omega_{CDM} \simeq 0.7$ and
$\Omega_{HDM} \simeq 0.3$. One can also get a flat universe
with $\Omega_{tot} = 1$ and low matter density if the
cosmological constant $\Lambda$ is nonzero. Such models
also give a satisfactory description of the observed
structure \cite{lam}.

The common shortcoming of these models is that they all demand a
certain amount of fine tuning. For example, one would generally
expect that the contribution from hot and cold dark matter to
differ by many orders of magnitude.

Recently there appeared a renewed interest \cite{ma1} to the
idea of structure formation with unstable particles \cite{GSV}.
The recent models assume that there exists a massive
long-lived particle, usually the tau neutrino with mass
in the MeV range, decaying into massless species at the
epoch when the mass density of the parent particles
dominates the energy density of the universe.
Correspondingly, the present-day energy density of
relativistic particles is bigger than in the standard
scenario and the onset of the MD stage takes place later.
All these models have so far been based on the assumption
of two unrelated components. A possible exception might be
the models with non-zero $\Lambda$ if a consistent adjustment
mechanism is found which, as argued in ref. \cite{ad1},
leads generically to an noncompensated amount of vacuum
energy of order $m_{Pl}^2 / t^2$.

In this letter we will consider a model for the unstable
particle scenario in which both the unstable particles
and the particles of cold (or possibly warm) dark matter
are closely inter-connected. In fact, the decay of the
former produces particles of the present-day dark matter.

A necessary background model of this kind in particle
physics was proposed some time ago in an attempt to find
a phenomenologically acceptable way to spontaneously break
R parity in supersymmetric extensions of the standard \21 model
\cite{MASI_pot3}. The breaking of R parity occurs around the
electroweak scale and implies the violation of lepton number
\cite{CMP}. Moreover, in the simplest realization adopted
in ref. \cite{MASI_pot3} lepton number is violated only
through R parity violating interactions. As a result,
the magnitude of the R parity violating effects is
related to the large Majorana type mass of the tau
neutrino, which could easily reach the MeV range.
As shown in ref. \cite{RPMSW} the \nt is unstable and
decays into the light \nm plus a majoron $J$ \cite{fae}.
A possible breaking of global symmetries by gravity would
give a nonzero mass to the corresponding (pseudo) goldstone
boson \cite{Goran}.  In this context it has been
suggested that the majoron may pick up a mass of the order
KeV, which is in the right range for it to play the
role of cold dark matter particle \cite{Venya}.

We assume that the values of mass and lifetime of \nt
are such that there existed a period
when it dominated the cosmological energy density.
We denote the corresponding red-shift when it started as $z_1$.
Another interesting moment was when the tau neutrino decayed
and the universe returned to an RD stage. The corresponding
red-shift is $z_d$. And at last there is the moment, which
is especially important for structure formation, when the
contemporary MD stage started. The onset of this stage was
at red-shift $z_2$.
The red-shift $z_1$ is determined by the equality of
energy densities of relativistic, $\rho_{rel}$, and
nonrelativistic, $\rho_m$, matter:
\beq{
z_1+1 \simeq 2.9 \times 10^8 r_\nu m_M (1 + 0.34 \xi^4 )^{-1}
\label{z1}
}
\eeq
Here  $m_M = m_{\nu_\tau} / \rm{MeV}$, $r_\nu$ is the ratio of
$\nu_\tau$ number density to that of normal massless neutrinos,
and $\xi = T_J / T_\gamma$.
For light \nt ($m<3$ MeV) $\xi \approx 1$ and for heavy \nt
($m>10$ MeV) $\xi \approx (4/11)^{1/3}$.
The value of the photon temperature at this moment is
$ T_1 \simeq  69 \rm{KeV} m_M r_\nu (1+ 0.34\xi^4)^{-1}$.
This result is valid if $T_1$ is larger than the majoron mass $m_J$.
We assume here that this is true.
The other two interesting red-shifts are given by
\beq{
z_d+1 \simeq 1.3 \times 10^5 (m_M r_\nu )^{-1/3} (\tau_y)^{-2/3}
\label{zd}
}\eeq
\beq{
z_2+1 \simeq 2.8 \times 10^4  \Omega h^2  (1 + 10^3 (m_M r_\nu)^{4/3}
\tau_y ^{2/3})^{-1}
\label{z2}
}\eeq

The upper bound on the universe age does not permit neutrinos to have
mass above a few tens eV or, in the case of nonzero cosmological
constant, a few hundreds eV \cite{zsk}. This bound can be
avoided only if the \nt is unstable or has annihilation
cross section considerably larger than in the standard model.
In fact in the scenario we consider here both ingredients are
present, so that neutrino masses in the MeV region are
cosmologically allowed.
The successful prediction for the nucleosynthesis of light elements
forbids Majorana tau neutrino masses in the range
$0.5 < m_M < 35 $ if the latter has the standard weak
annihilation cross section and is stable on the nucleosynthesis time
scale i.e $\tau > 100$ sec \cite{ktcs,BBNUTAU}. In our case, however,
this bound can be avoided because $\nu_\tau$ has a rather strong
diagonal Yukawa coupling to majorons,
\beq{
\cal{ L}_{diag} = g J \nu_\tau^T \sigma_2 \nu_\tau + h.c.
\label{ldiag}
}\eeq
In many majoron models $g = m_\nu /V$ where $V$ is the scale
at which spontaneous lepton number violation occurs which is,
by assumption, the weak scale, so that
$g = 10^{-5} m_M / (V/100\,\rm{GeV})$.
In order to suppress $r_\nu$ sufficiently so that the
massive $\nu_\tau$ would not distort the nucleosynthesis
predictions on the one hand, and sufficiently large for a
noticeable delay of the onset of the last MD stage on
the other, we need $g$ slightly above $10^{-4}$. This
value fits nicely with the assumption that the \nt mass
lies in the MeV range. It is also surprisingly close to the
values assumed in the spontaneously broken R parity model
\cite{RPMSW}. For such $g$ values \nt \nt annihilations
into majorons substantially reduces the $\nu_\tau$ number
density during nucleosynthesis and after
\footnote{A more detailed analysis of the nucleosynthesis
bounds on \nt masses and annihilation cross sections will
be presented elsewhere \cite{drpv}}.

Another useful restriction follows from the bound on the
total energy density in the universe. For any form of
matter we have an upper bound:
$\rho_x < \rho_{tot} = 10.5 \Omega h^2 \rm{KeV/cm^3}$.
Applying this bound to the massive majorons gives
$r_J < 0.1 \Omega h^2 /m_K$ where $m_K$ is the majoron
mass in KeV and $r_J = n_J / n_0$, where $n_J$ is the majoron
number density and $n_0$ is the number density of normal
massless neutrinos, $n_0 = 0.18 T_\nu^3 $.
The majorons in our model are produced in the early universe
either thermally or possibly by the lepton number violation
phase transition. In addition, they can be produced as a
result of $\nu_\tau$ decay at a relatively late stage.
For the latter $r_J^{decay} = r_\nu$ and for the former
one should naively expect $r_J^{therm} = (1/2)(4/3) = 2/3$.
If this is true the majoron mass should be too small to be
interesting for the process of structure formation, since
the characteristic scale of the structures to form first
would be too large, as in the hot neutrino scenario.
However, it is important to note that the majorons are
rather strongly self-coupled particles, as they possess
an interaction $\lambda J^4$ whose coupling constant
$\lambda$ could be as large as $10^{-2}$. This interaction
in the second order leads to the process of {\sl cannibalism},
when collisions of four J's produce only two. The freeze-out
of species whose number density was reduced by multiparticle
collisions $n \rightarrow 2$ was considered in refs. \cite{dadth,cmh}
in a simplified approach. Here we have numerically integrated the
kinetic equation governing this process, under the usual assumptions
that majorons are in kinetic equilibrium and described by Boltzman
statistics. Then in complete analogy to two-body annihilation
(see e.g. \cite{gg}) we get
\beq{
{dr_J \over dt} = { N r_J^2 (r_J^{(eq)2} - r_J^2)
\over 2^5\pi^4 n_0 r_J^{(eq)4} }
\int^\infty_{16m_J^2} ds \sigma (s) s(s-4m_J^2) {T_J \over \sqrt s}
K_1 \left( {T_J \over \sqrt s} \right)
\label{rdot}
}\eeq
where $N$ is the combinatorial factor related to identical $J$'s
(we take $N=2$), $K_1$ is the McDonalds (Bessel) function,
$r_R = n_J / n_0$, $r_J^{eq} = n_J^{eq} / n_0$, where
$n_J^{eq}$ is the equilibrium number density of majorons
given by the Boltzman distribution with temperature $T_J$.

It is a rather tedious job to calculate the cross section of the
process $ 4J \rightarrow 2J$ but, since there is considerable
freedom in the choice of the value of $\lambda$ which is not
known anyhow, we make a simple estimate of the cross section
$\sigma (s)$ assuming that the amplitude of the process is a
constant, $A = \lambda^2 /m_J^2$. Since the photon temperature
of is not equal to the temperature of the majorons, we need
another equation which is provided by the covariant law of
majoron energy density conservation.

In our numerical estimates we have assumed the following
parameter values:
$r_\nu m_M = 10^{-2}$, $m_M = 10$,
$\tau_\nu  = 5$ yr., and $\Omega h^2 = 0.5$.
These numerical values are only illustrative, we use them
as a reference set in order to show that the model can be
self-consistent. With these numbers we get $T_1 \approx 630$ eV,
$T_d \approx 50$ eV, and $T_2 \approx  0.45$ eV. The frozen number
density of majorons is found to be sufficiently small,
$r_J \approx 0.2$.
We assume that the majoron mass is $m_K = 0.3$. Correspondingly
the characteristic size of the structures first formed can be two
orders of magnitude smaller than in the neutrino universe,
i.e. M = a few $\times 10^{12} M_{\odot}$.

Some bounds on the properties of MeV tau neutrinos and majorons
can be found from astrophysics. In order to avoid excessive stellar
cooling through majoron emission their mean free path inside a star
($l^{free}_J$) should be smaller than the radius of the latter. For
a supernova with temperature $T \approx 10$ MeV the latter is
$l^{free}_J = 1/\sigma n_\nu = 10^{10} (10^{-5}/g)^4 $cm. For
$g > 10^{-4}$ we are safe. Another possible restriction arises
from the observed neutrino flux from SN87A. There could be delayed
neutrinos originated from the decay $\nu_\tau \rightarrow J+\nu$
\cite{Soares}. But, for example, in the model of ref. \cite{MASI_pot3}
the main \nt decay channel is into \nm so that this restriction is
also harmless.

By assumption the majoron arises from the spontaneous
violation of the global U(1) lepton number symmetry
which generates the neutrino masses. For \nt masses
in the MeV range the majoron has a sizeable diagonal
Yukawa coupling to the \nt of order $g=10^{-4}$.
The majoron also has a much weaker nondiagonal
coupling to muon neutrinos which induces the decay
$\nu_\tau \rightarrow \nu_\mu + J$ with lifetime of
order of years. In order to avoid the supernova bound
we assume the nondiagonal coupling to \ne to be negligible.
Moreover the diagonal couplings of the majoron to \ne and
\nm are severely restricted in order to ensure that the
majoron decay lifetime $J \ra \nu \nu$ is larger than
the age of the universe, i.e. $g_{\nu \nu} \lsim 10^{-17}$.
Using the estimate $m_{\nu} = g_{\nu \nu} V$ and a typical
value for the lepton number violation scale around
$10^3$ GeV we obtain that the \ne and \nm masses
will be around $10^{-5}$ eV, i.e., just the range
where the solar \neu deficit will be explained by
long wavelength or just-so \ne to \nm oscillations.

Moreover, our unstable dark matter majoron will have
tiny loop-induced decays to two photons $J \ra \gamma \gamma$
which will lead to the existence of an X-ray line at
$E_\gamma = m_J/2$ which could be detectable
\cite{Venya}.

The self-coupling $\lambda J^4$ between majorons with
$\lambda$ in the range from $10^{-2}$ to $10^{-4}$
leads to an interaction cross section
$\sigma_{el} = \lambda^2 /(64 \pi m_J^2)$ which will
maintain kinetic equilibrium between majorons till the
present day. The corresponding majoron mean free path
$l_{free}=(\sigma_{el} n_J)^{-1}$ with
$n_J = n_{cosm} = 10 \Omega h^2 m_K^{-1}/ \rm{cm}^3$
is about 100 Kpc$(m_K /\lambda_3^2)$ where
$m_K = m_J / \rm{KeV}$ and $\lambda_3 = 10^3\lambda$.
Inside galaxies with
$n_J = 3 \times 10^4 \rm{cm}^{-3} m_K^{-1}$
this mean free path is much smaller than the galactic
size, $l_{free} = 7(m_K/\lambda_3^2)$ pc.

The possibility that dark matter can have
a relatively strong self-interaction, while
being essentially decoupled from usual matter,
has been considered earlier \cite{cmh,m}. The
behaviour of such matter is intermediate
between that found in the CDM and HDM models.
Structure formation with self-interacting dark
matter particles has been recently criticized
\cite{lss}. Our scheme however is different
from the simple original version and avoids
at least some of these criticisms. The main
objection of ref. \cite{lss} is that it is
impossible to generate enough small scale power
to account for the damped Ly$\alpha$ systems
without producing too much power at cluster
scales. Any model with decaying particles and
early MD stage, such as ours, automatically takes
care of this, since it has a suppressed power at
cluster scales and simultaneously a large power on
smaller scales due to rising fluctuations at the earlier
MD stage. Another point raised in ref \cite{lss} is that the galaxy
merging should be quite different from what is expected in the usual
collisionless dark matter case, and that the motion of galaxies
with respect to the cosmic background would result in the stripping
of the dark matter halos from galaxies. However, these points deserve
further investigation. For example, the stripping of the halos may
not take place if all the dark matter is clustered inside the halos
and the cosmological background is absent. Present data show that
the galactic halos extend up to two or more hundred kiloparsecs,
and the 100\% clustering at a large scale could account for
$\Omega \approx 1$. Note also that the process of clustering
of self-interacting matter differs from that of collisionless
or baryonic matter because the former does not loose energy
so easily by radiating photons.

The self-interacting dark matter would form an isothermal
matter distribution in the halos, and since the self-gravitating
isothermal gas has a mass density which goes down with distance
as $1/r^2$, it gives a natural explanation of the observed
flat rotational curves. It was recently noted \cite{rotcur} that the
shape of the rotation curves inside or not far from the luminous
center (for dwarves and also possibly for spirals)
does not agree with the assumption of the collisionless dark matter.
While this statement depends on how matter is distributed in
galaxies, if confirmed, it may be a strong argument in favor
of self-interacting dark matter. Another test of the model
(suggested to us by S. White) is the shape of the halo:
for self-interacting dark matter it should be more spherical
than for the sterile one.

Last but not least, the scenario described here can be implemented
rather naturally in majoron extensions of the standard
model of particle physics. As an example, supersymmetry
with spontaneously broken R parity provides a consistent
model for our scenario which incorporates all the
necessary ingredients.

{\bf Acknowledgements}

This work was supported by DGICYT under grants
PB92-0084 and SAB94-0089 (A. D.). S.P. was supported by
Conselleria de Educaci\'o i Ci\`encia of Generalitat Valenciana.

\end{document}